\documentclass[apjl]{emulateapj}
\def\ergs{ergs~s$^{-1}$}
\def\ergcms{ergs$^{-1}$~cm$^{-2}$~s$^{-1}$}

\def\hst{{\it HST}}
\def\cha{{\it Chandra}}
\def\xmm{{\it XMM}}
\def\asca{{\it ASCA}}

\begin{document}

\title{Optical Counterpart of the Ultraluminous X-ray Source IC~342~X-1}

\author{Hua Feng\altaffilmark{1} and Philip Kaaret}

\affil{Department of Physics and Astronomy, The University of Iowa, Van
Allen Hall, Iowa City, IA 52242}
\altaffiltext{1}{Email: hua-feng@uiowa.edu}

\shortauthors{Feng and Kaaret}
\shorttitle{Optical Counterpart of IC 342 X-1}

\begin{abstract}
We present \cha\ and \hst\ observations of the ultraluminous X-ray source (ULX) IC 342 X-1. The \cha\ and \hst\ images are aligned using two  X-ray emitting foreground stars. The astrometry corrected position for X-1 is $\rm R.A. = 03^h45^m55\fs61$, $\rm Decl. = +68\arcdeg04\arcmin55\farcs3$ (J2000) with an error circle of $0.2\arcsec$.  One extended optical source is found in the error circle, which could be the optical counterpart of X-1.  The source shows an extended feature in \hst\ images at long wavelengths, which is likely to be a superposition of two point sources, although it is possible that the dimmer one could be a jet.  Both sources are much redder than typical for ULX optical counterparts.  The brighter one has an absolute magnitude $M_V=-5.2\pm0.2$ and $(B-V)_0 = 0.66 \pm 0.13$ and the dimmer star is not detected in $B$ and has $(B-V)_0 > 2.1$.  Their colors are consistent with an F8 to G0 Ib supergiant or a carbon star, respectively.  However, it is likely that part or most of the optical emission may be due to X-rays reprocessed by the companion star or the accretion disk.  The stellar neighborhood of IC 342 X-1 lacks O stars and has a minimum age of $\sim 10$~Myr.  This excludes the possibility that the surrounding nebula is powered by an energetic explosion of a single massive star that formed a black hole. We suggest that the nebula is most likely powered by an outflow from the X-ray source.
\end{abstract}


\section{Introduction}

Ultraluminous X-ray sources, the brightest nonnuclear X-ray point sources in external galaxies, still hide their nature from us.  Their luminosity of $3\times10^{39}$ to $10^{41}$~\ergs, inferred assuming isotropic emission, is too high to be powered by accretion onto stellar mass black holes that are formed in the collapse of normal stars with normal metallicity, and thus may suggest the presence of intermediate mass black holes \citep[IMBHs;][]{col99,kaa01}.  However, if the emission is beamed \citep{kin01,kor02} or truly super-Eddington \citep{wat01,beg02}, IMBHs are not required.  

Multiwavelength observations have provided powerful means to determine the nature of ULXs.  \citet{pak02} carried out an optical survey of bright ULXs and discovered that several ULXs are associated with optical nebulae.  These nebulae have very large spatial scales of a few hundred parsecs, and present line emission. In particular, He~{\sc ii} $\lambda$4686 emission line was detected in the nebula surrounding Holmberg II X-1, and was used to determine that the flux of reprocessed X-rays from the central ULX was at least $4\times10^{39}$~\ergs\ \citep{kaa04}.  Identification of an optical counterpart of a ULX has been accomplished in several cases. However, it is hard to determine whether or not the optical light from the companion star is contaminated by the outer region of the accretion disk, thus stellar classification based on optical colors may not be reliable.  The far-ultraviolet spectrum of NGC 5204 X-1 suggests a B0 Ib companion, but contains some peculiar emission lines \citep{liu04}.  Identification of a giant branch companion has been made for the ULX X41.4+60 in M82 based on the detection of a 62~day X-ray modulation interpreted as the orbital period \citep{kaa06a,kaa06b,kaa07}.  Identification of the companion stars for more ULXs would enable us to place constraints on the binary evolution of these systems.

IC 432 is a nearly face-on starburst spiral galaxy.  IC~342 X-1 was discovered with {\it Einstein} \citep{fab87} and confirmed as a point source with {\it ROSAT} observations \citep{bre93}.  Using \asca\ observations in 1993, \citet{oka98} found significant variability on time scales of hours from X-1, suggesting it was a compact object.  The source changed its spectrum dramatically in a 2000 \asca\ observation \citep{kub01}, similar to the spectral state transitions found in Galactic black hole binaries.  A possible X-ray periodicity of 31 or 41 hours from X-1 was also detected in the same \asca\ observation in 2000 \citep{sug01}.  X-1 is associated with a ``tooth''-shaped optical nebula \citep{pak02} with emission lines similar to a supernova remnant \citep[SNR;][]{pak02,rob03,abo07}.  \citet{rob03} estimated that the initial energy required to power the nebula was at least $2\times10^{51}$~ergs, greater than a canonical SNR.  Also, \citet{rob03} found two regions of [O~{\sc iii}] emission located oppositely on the shell of the nebula, which could be X-ray photoionized and thus indicate anisotropic emission from X-1. Instead, \citet{gri06} found that the [O~{\sc iii}] emission closely follows the H$\alpha$ emission and is smoothly distributed in the nebula, implying it is shock excited. \citet{abo07} found evidence for photoionization in the central regions of the nebula.

To search for an optical counterpart of IC~342 X-1, we obtained \cha\ and \hst\ observations. The observations and data are described in \S~\ref{sec:obs}.  The \cha\ and \hst\ images are aligned by matching two foreground stars in both images (\S~\ref{sec:astr}), which leads to the identification of a unique optical counterpart of X-1 (\S~\ref{sec:cp}). The photometry for the optical counterpart of X-1 and its nearby starforming region is performed in \S~\ref{sec:photo}.  A multiwavelength spectrum from optical to X-ray is presented in \S~\ref{sec:multi}. An emission line image obtained using \hst\ of the tooth nebula is shown in \S~\ref{sec:halpha}. All results are discussed in \S~\ref{sec:diss}.  We adopt a distance to IC~342 of $3.3 \pm 0.3$~Mpc, which is estimated from observations of 20 Cepheids in the galaxy \citep{sah02}.

\section{Observations and Analysis}
\label{sec:obs}

\begin{figure*}
\centering
\includegraphics[width=1\columnwidth]{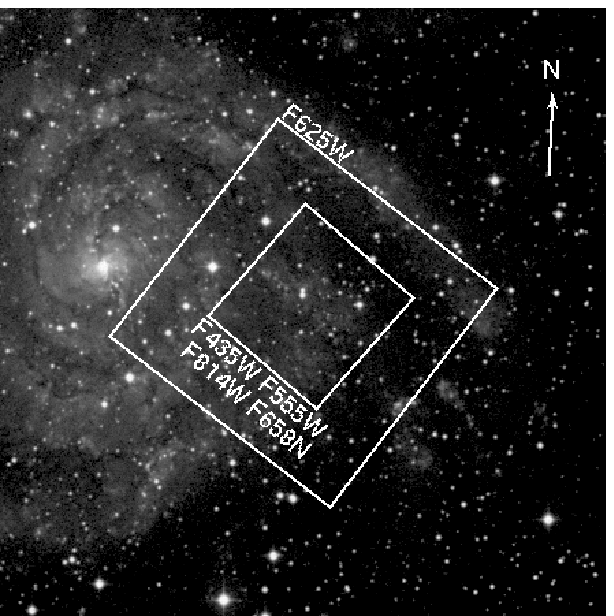}\\
\includegraphics[width=1\columnwidth]{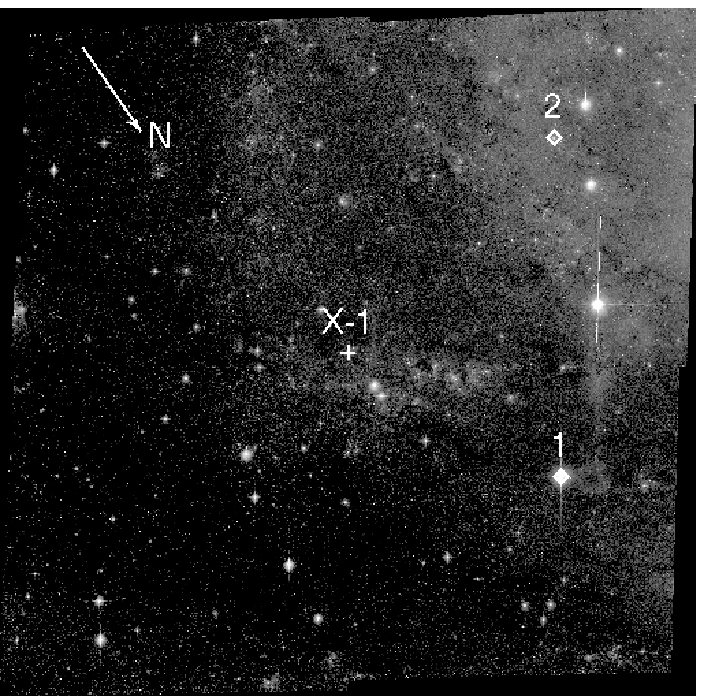}
\caption{
{\it Top}: DSS blue band image of IC 342. The large rectangle indicates the field of view of the mosaic F625W image, and the small one indicates the field of view for the F435W, F555W, F814W, and F658N images. The arrow points north and has a length of 2\arcmin.
{\it Bottom}: Mosaic F625W image. The positions of X-1 and the two sources for astrometry are labeled. The arrow points north and has a length of 1\arcmin.
\label{fig:1010}}
\end{figure*}

\hst\ observed the IC~342 X-1 region with the Advanced Camera for Surveys (ACS) under the proposal 10768 (PI: P. Kaaret) in December 2005. Four observations, with an exposure of 920~s each, were made using the Wide Field Camera (WFC) with a broadband filter F625W (Sloan Digital Sky Survey R). Each observation had the same orientation, but had IC~342 X-1 located in a different corner of the image in order to create a large mosaic image around X-1.  Another four WFC images were obtained with X-1 at the center using three broadband filters, F435W (Johnson B), F555W (Johnson V), F814W (Broad I), and a narrowband filter, F658N (H$\alpha$ + [N~{\sc ii}]). Each of the three broadband images has an exposure of 1800~s, and the narrowband one has an exposure of 2400~s.  The fields of view for all images are shown on an STScI Digitized Sky Survey (DSS) blue image of IC~342 (Figure~\ref{fig:1010} {\it top}).  We used the standard drizzled ACS images for all analysis.

\begin{deluxetable*}{cllcllccc}
\tablewidth{0pc}
\tablecolumns{9}
\tablecaption{\hst\ and \cha\ positions of the two sources for astrometry correction.
\label{tab:cp}}
\tablehead{
\colhead{} & \multicolumn{2}{c}{\hst} & \colhead{} & \multicolumn{5}{c}{\cha}\\
\cline{2-3} \cline{5-9}\\
\colhead{No.} & \colhead{R.A.} & \colhead{Decl.} & \colhead{} & \colhead{R.A.} & \colhead{Decl.} & \colhead{$\delta$} & \colhead{$\Delta_{\rm RA}$} & \colhead{$\Delta_{\rm Decl}$} \\
\colhead{} & \colhead{(J2000)} & \colhead{(J2000)} & \colhead{} &  \colhead{(J2000)} & \colhead{(J2000)} & \colhead{(\arcsec)} & \colhead{(\arcsec)} & \colhead{(\arcsec)}\\
\colhead{(1)} & \colhead{(2)} & \colhead{(3)} & \colhead{} &  \colhead{(4)} & \colhead{(5)} & \colhead{(6)} & \colhead{(7)} & \colhead{(8)}
}
\startdata
1 & 03 46 06.592 & +68 07 05.42 && 03 46 06.627 & +68 07 05.49 & 0.09 & 0.00 & 0.07 \\
2 & 03 46 25.985 & +68 04 19.87 && 03 46 26.004 & +68 04 19.94 & 0.41 & 0.24 & 0.06 \\
\enddata

\tablecomments{$\delta$ is the statistical 90\% error radius for \cha\ positions. $\Delta$ is the systematic offset of the {\tt wavdetect} position caused by the asymmetry of the point spread function.}
\end{deluxetable*}

\begin{figure*}
\centering
\includegraphics[width=0.31\textwidth]{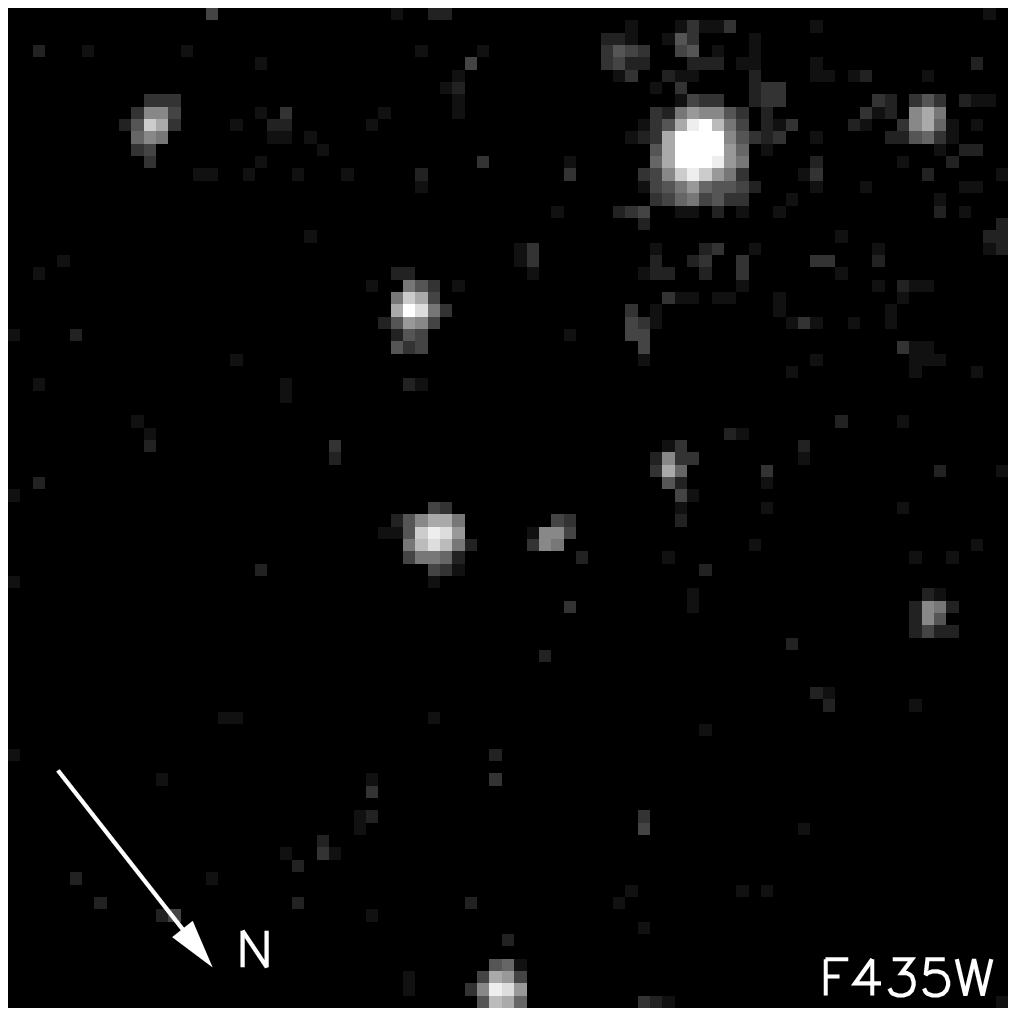}
\includegraphics[width=0.31\textwidth]{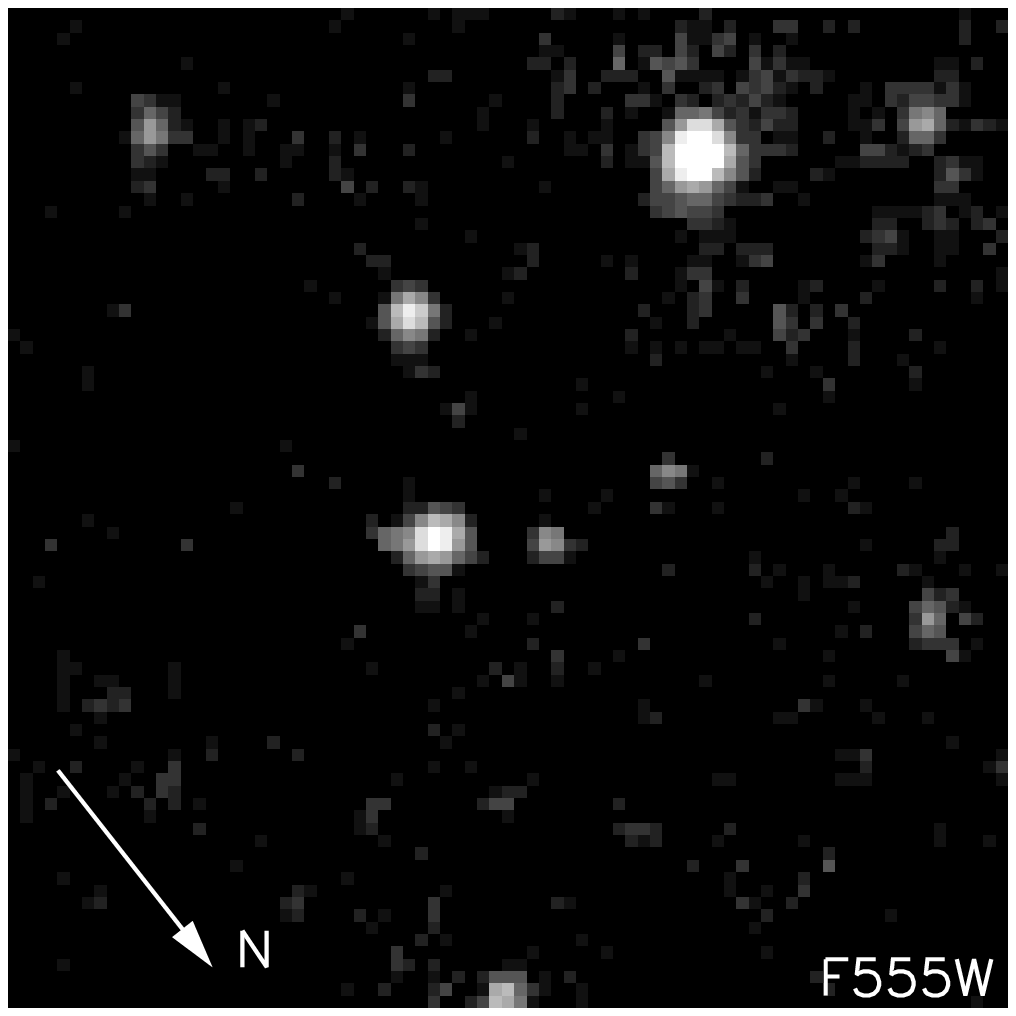}
\includegraphics[width=0.31\textwidth]{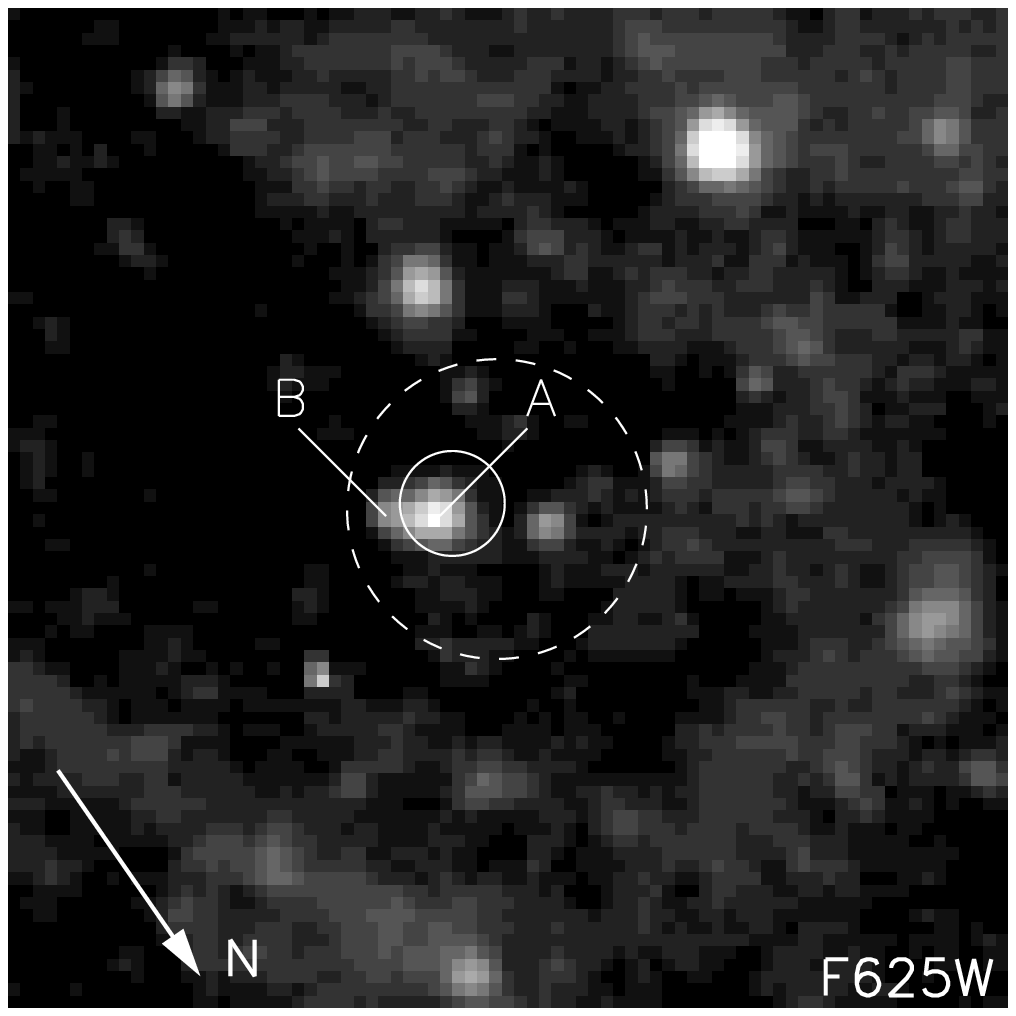}\\
\includegraphics[width=0.31\textwidth]{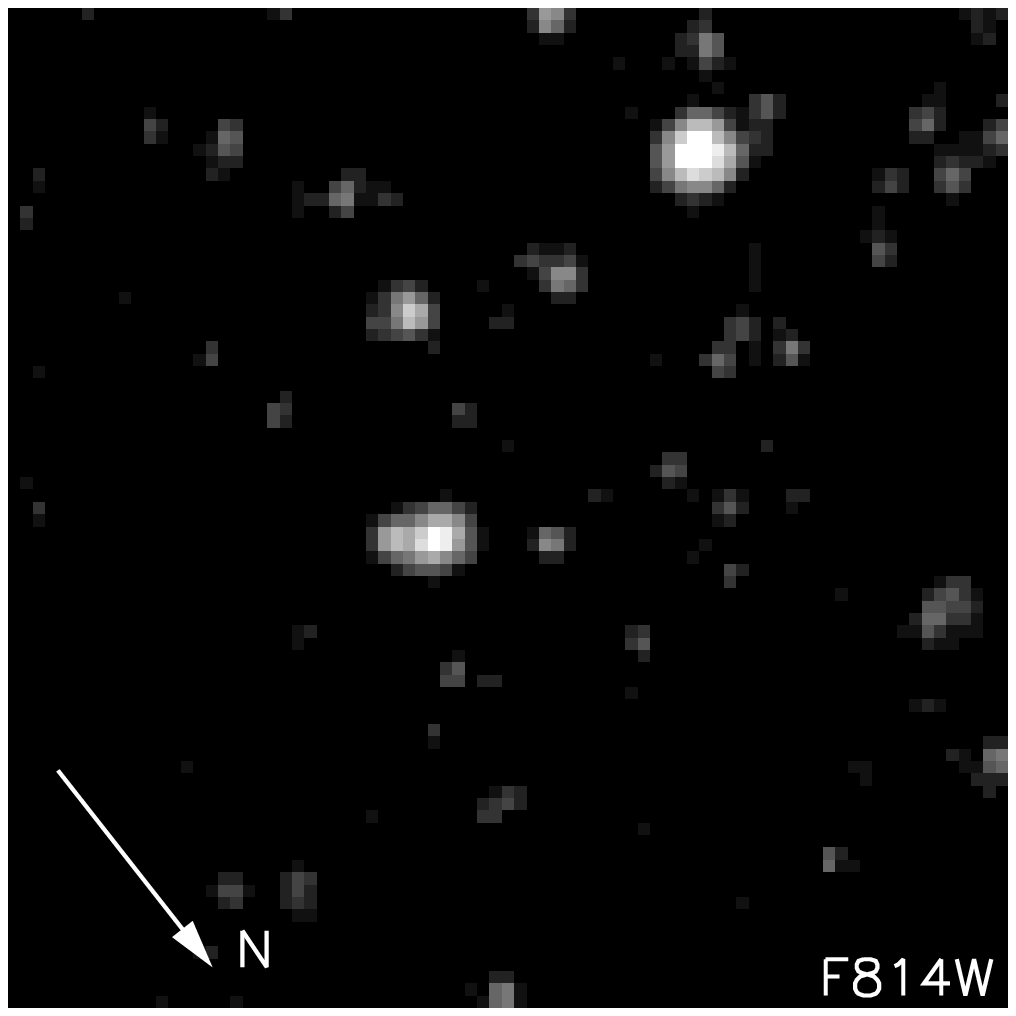}
\includegraphics[width=0.31\textwidth]{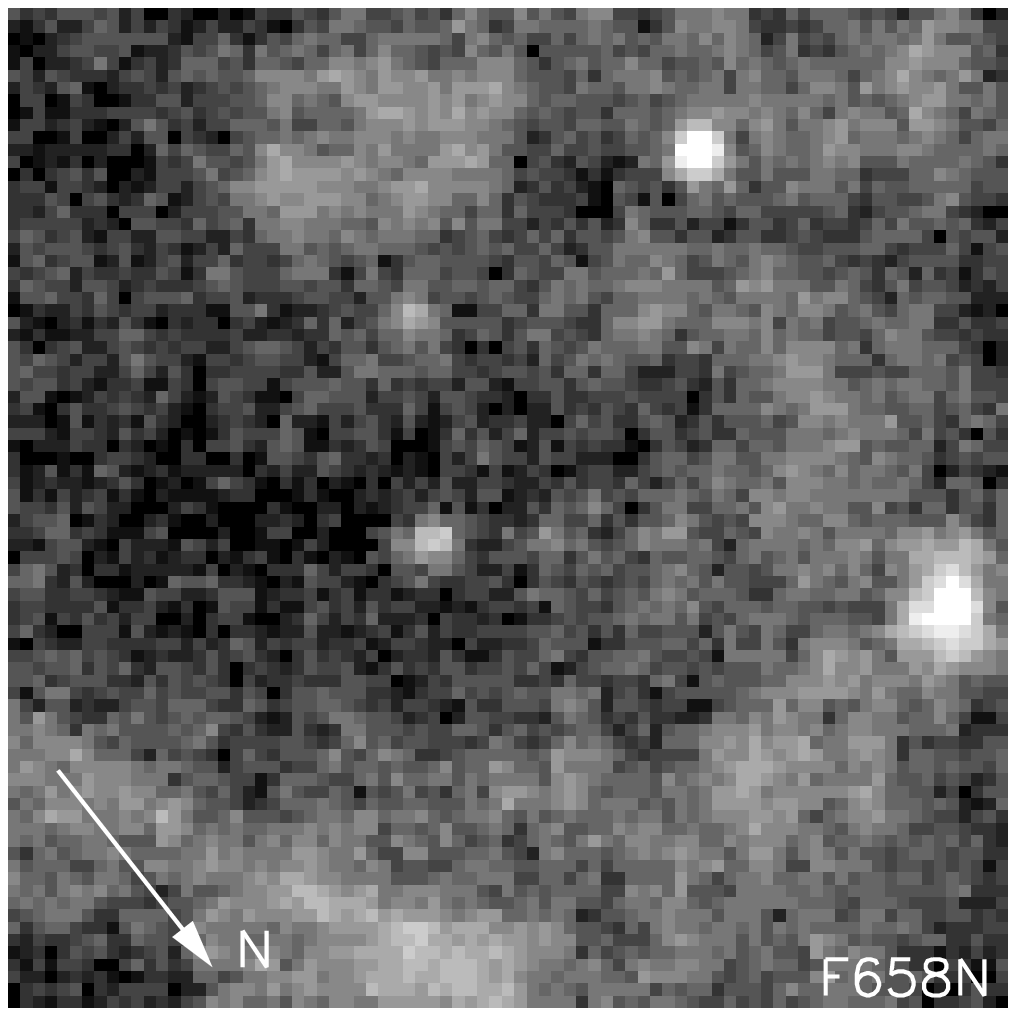}
\caption{
Images in the region around IC~342 X-1. In the F625W image, the dashed circle indicates the directly measured \cha\ position of X-1 with an uncertainty of $0.6\arcsec$, the solid circle indicates the corrected X-1 position with an uncertainty of $0.21\arcsec$, and the two possible optical counterparts, stars A and B, are labeled.  The arrows points north and are 1\arcsec\ long.
\label{fig:x1}}
\end{figure*}

\begin{figure*}
\centering
\includegraphics[width=0.35\textwidth]{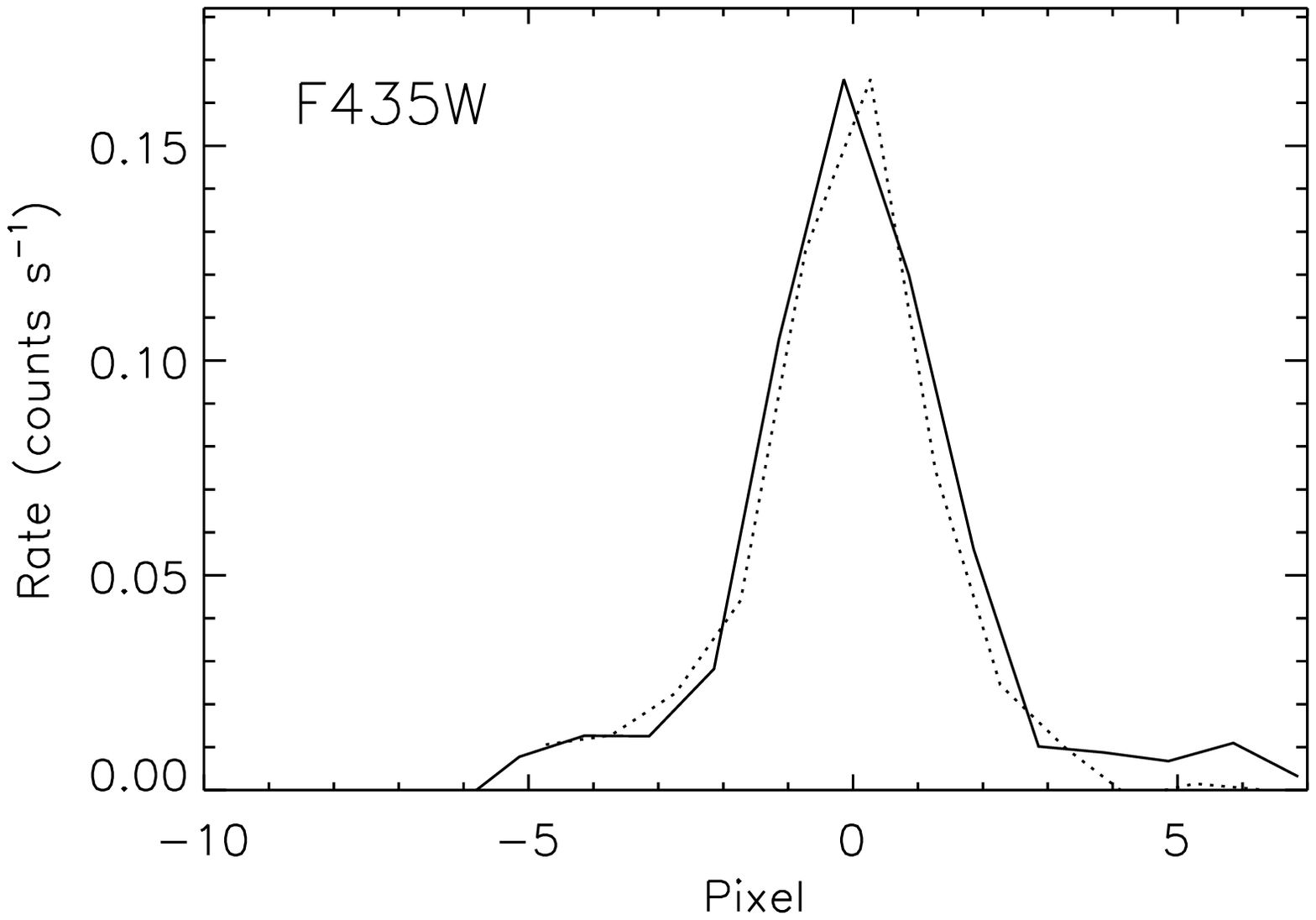}\hspace{0.1\textwidth}
\includegraphics[width=0.35\textwidth]{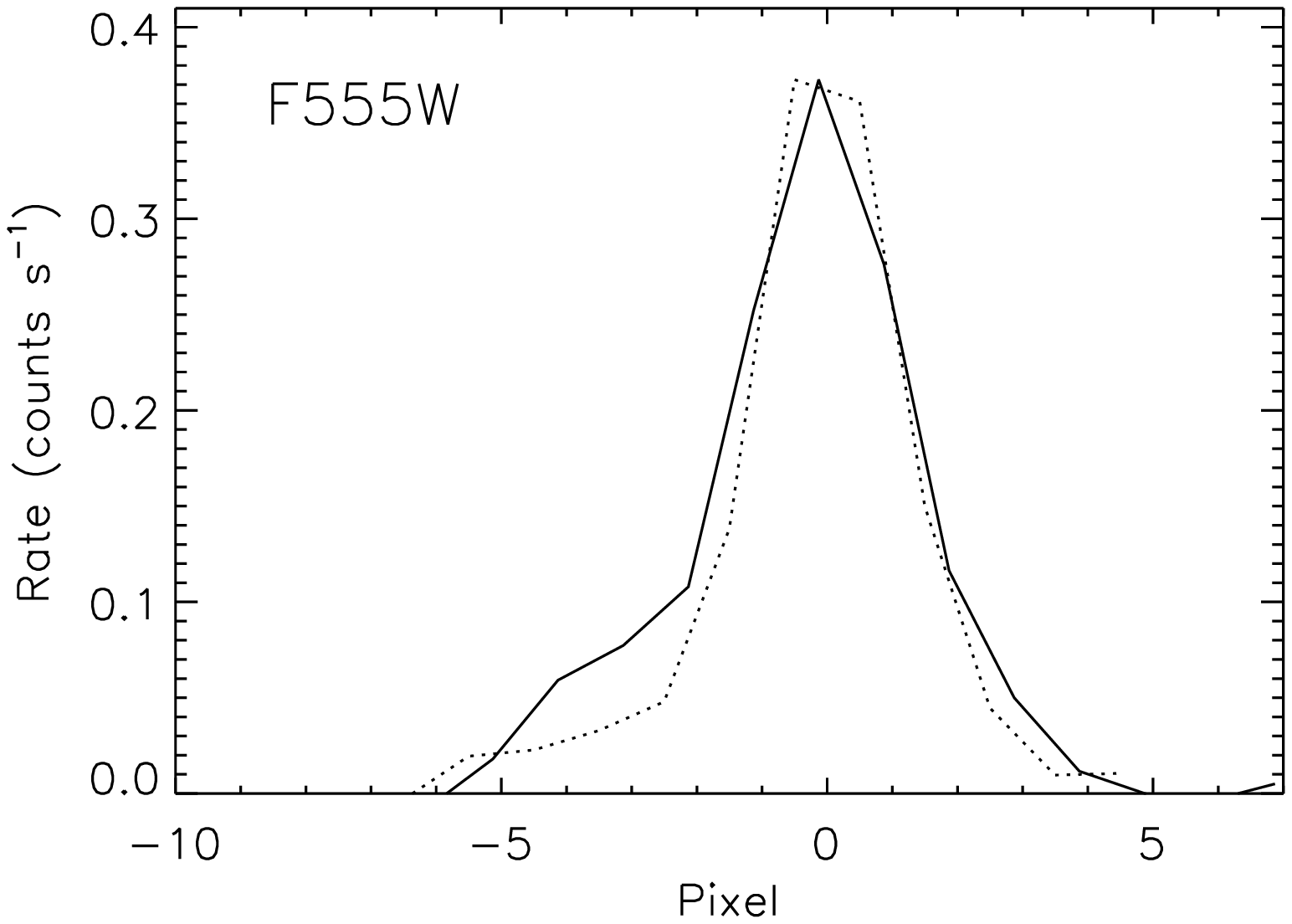}\\
\includegraphics[width=0.35\textwidth]{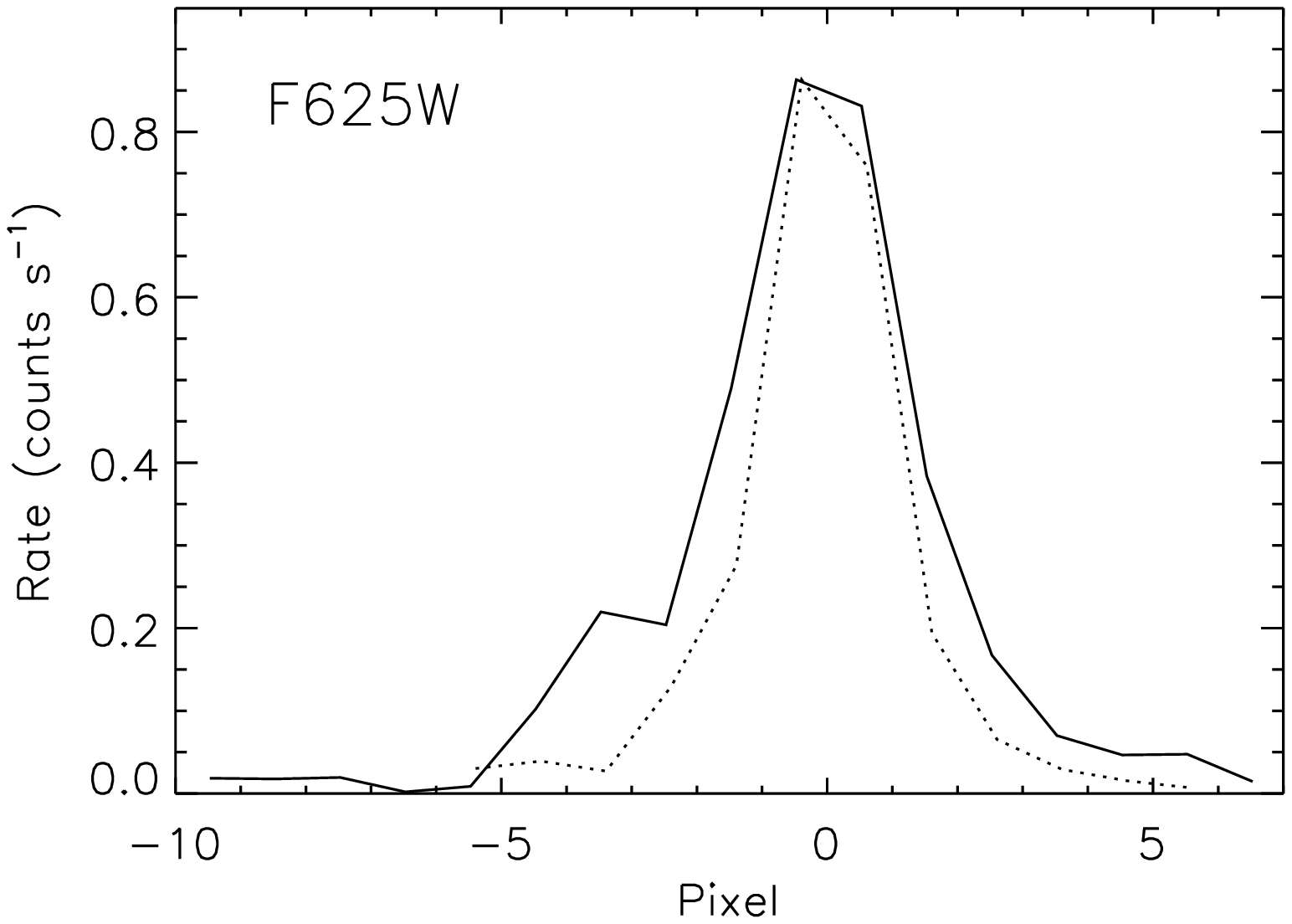}\hspace{0.1\textwidth}
\includegraphics[width=0.35\textwidth]{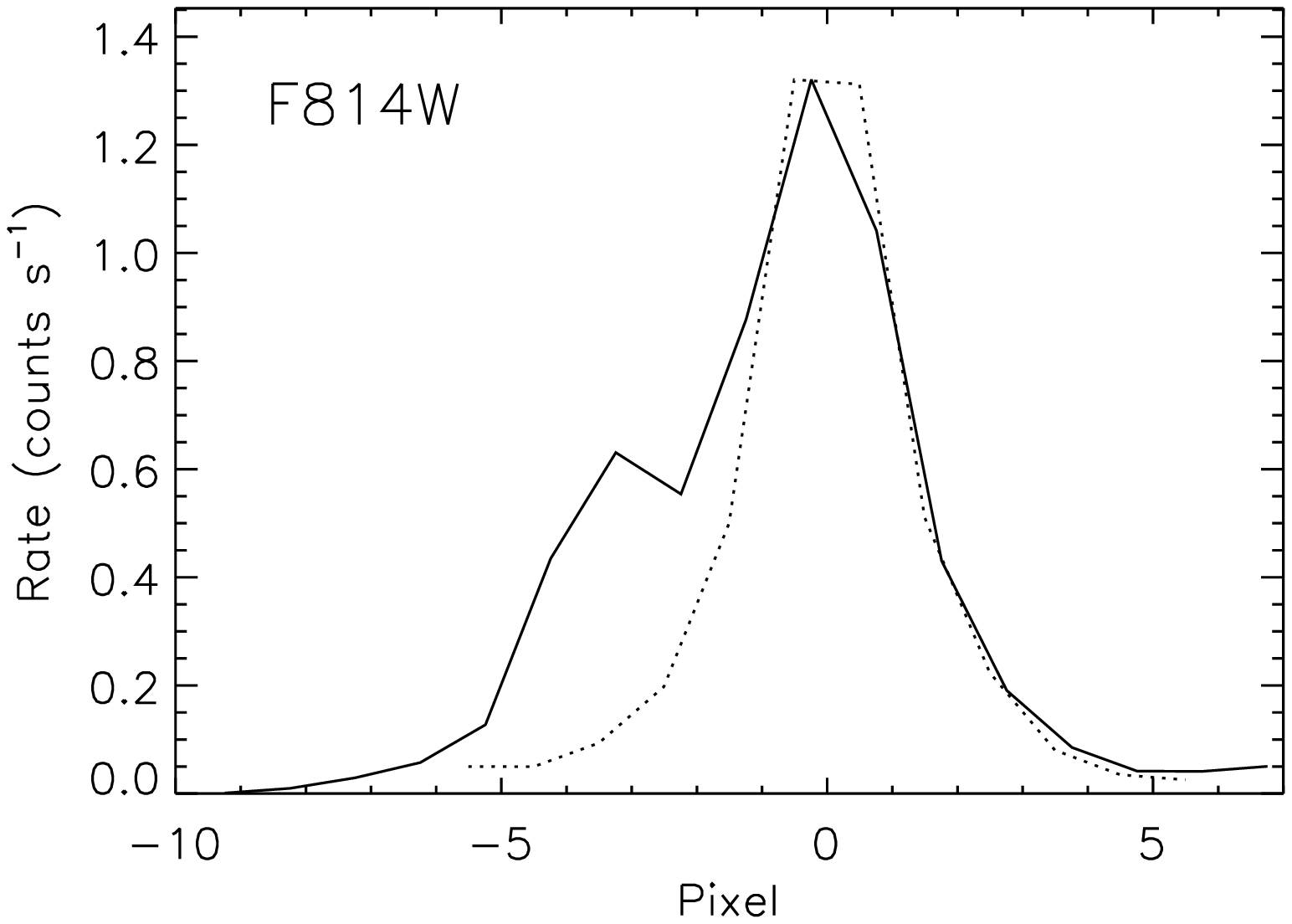}
\caption{
Profiles of brightness along the X ({\it solid line}) and Y ({\it dotted line}) directions of the CCD across the brightest pixel of star A.  X and Y profiles are centered at their best-fit centroids. There is a secondary component on the X profile at long wavelengths, corresponding to star B in Figure~\ref{fig:x1}.
\label{fig:profile}}
\end{figure*}

The \cha\ X-ray Observatory observed IC~342 X-1 on 2006 Nov 12 (ObsID 7069; PI: P. Kaaret) with the Advanced CCD Imaging Spectrometer (ACIS). An X-ray image was created in the 0.3-8 keV range, covering the field of view of the mosaic F625W image. We performed source detection using the {\tt wavdetect} tool in CIAO 3.3.0.1 applying an exposure map calculated for a power-law spectrum with a photon index $\Gamma=1.5$  and Galactic absorption $N_{\rm H} = 3.02 \times 10^{21}$~cm$^{-2}$.

\section{Astrometry Correction}
\label{sec:astr}

Each \hst\ image was aligned to the sources in the 2MASS catalog \citep{skr06} using the Graphical Astronomy and Image Analysis Tool (GAIA) to correct the absolute astrometry.  We selected catalog stars that appear as pointlike sources in an \hst\ image, and found 22 to 35 counterparts in each image.  The rms (root mean square) deviation of the \hst\ source positions versus the 2MASS catalog positions is $\la$2 pixels ($\approx 0.1\arcsec$).  The astrometric accuracy for 2MASS stars with $13<K_S<15$ (ones used here) has an rms of around $0.1\arcsec$ along the right ascension \citep[see Fig.~20 in][]{skr06}, which corresponds to a 90\% positional error of $0.22\arcsec$ (assuming a Rayleigh distribution). After correction, the statistical accuracy of the alignment of each \hst\ image relative to the 2MASS frame is about $\delta_{\rm H}=0.22\arcsec/\sqrt{N}\approx0.04\arcsec$ at 90\% confidence, where $N$ (= 22 to 35) is the number of counterparts.

The four F625W images, after astrometry correction, are combined into a mosaic image (see Figure~\ref{fig:1010} {\it bottom}) using the IRAF package MSCRED. A flux offset was imposed on each image in order to equalize the mean flux in the overlapping image regions.  This helps remove image artifacts at the image boundaries.

Six counterparts are found between the X-ray image and the F625W mosaic image. Two of them are located on a different ACIS CCD chip from X-1, thus not suitable for astrometry.  Another two sources have only 15 and 17 photons detected by \cha, and the brightest pixel around the source region contains only 2 and 4 photons, respectively; we decided not to use them for astrometry due to their low significance.  The other two sources have 617 and 74 photons detected by \cha, respectively, with a significance of 177$\sigma$ and 18$\sigma$ given by {\tt wavdetect}.  These two sources are used for astrometry, and their \hst\ and \cha\ positions are shown in Table~\ref{tab:cp}.  Source 1 is highly saturated in the \hst\ image, and source 2 is moderately saturated. Their \hst\ positions are obtained by measuring the crosspoint of the diffraction spikes, and the errors are estimated to be less than 0.05\arcsec.  The X-ray positions and statistical errors are obtained from {\tt wavdetect}. There are systematic offsets in the positions reported by {\tt wavdetect} caused by the shape of the point spread function.  To estimate these, we performed simulations using MARX of sources located at the observed positions, fit the simulated images using {\tt wavdetect}, and calculated the offset between the sky position input to the simulation and the position reported by {\tt wavdetect}.  These systematic offsets are listed in column (7) and (8) of Table~\ref{tab:cp} for each source. The corrected \cha\ positions can be obtained by adding the positions quoted in Table~\ref{tab:cp} and the corresponding offset.  Source 1 is referred as source 11 in \citet{bau03} and X12 in \citet{kon03}.  It appears to be a foreground star.  Source 2 appears in several catalogs and its colors and X-ray to optical ratio \citep[$\ga 0.001$; as defined in][]{mac82} are also consistent with a foreground star.

From {\tt wavdetect}, the position of X-1 is obtained as $\rm R.A. = 03^h45^m55\fs636$, $\rm Decl. = +68\arcdeg04\arcmin55\farcs41$ (J2000) with a statistical error of $\delta_{\rm X1}=0.02\arcsec$ (90\% confidence).  The system offset in the {\tt wavdetect} position estimated using a MARX simulation as described above is $\Delta_{\rm RA} = 0.03\arcsec$, $\Delta_{\rm Dec} = 0.01\arcsec$.  The absolute astrometric error for \cha\ is $\delta_{\rm C}=0.6\arcsec$ at 90\% confidence\footnote{\url{http://cxc.harvard.edu/cal/ASPECT/celmon/}}.

\begin{deluxetable*}{cccccc}[t]
\tablewidth{0pc}
\tablecolumns{6}
\tablecaption{Magnitudes and colors of the two stars near IC~342 X-1
\label{tab:x1cp}}
\tablehead{
\colhead{} & \multicolumn{2}{c}{Star A} & \colhead{} & \multicolumn{2}{c}{Star B} \\
\cline{2-3} \cline{5-6}
\colhead{Filter/Band} & \colhead{$m$} & \colhead{$m_0$} && \colhead{$m$} & \colhead{$m_0$} \\
\colhead{} & \colhead{(mag)} & \colhead{(mag)} & \colhead{} & \colhead{(mag)} & \colhead{(mag)}
}
\startdata
\multicolumn{6}{c}{ST magnitudes} \\
\noalign{\vskip 0.8ex}\hline\noalign{\vskip 0.8ex}
F435W & 24.68 $\pm$ 0.07 & 22.40 $\pm$ 0.09 && $>$ 28.1 & $>$ 25.8 \\
F555W & 24.15 $\pm$ 0.06 & 22.33 $\pm$ 0.07 && 26.08 $\pm$ 0.17 & 24.30 $\pm$ 0.18 \\
F625W & 23.95 $\pm$ 0.06 & 22.48 $\pm$ 0.07 && 25.44 $\pm$ 0.13 & 24.02 $\pm$ 0.15 \\
F814W & 23.77 $\pm$ 0.05 & 22.68 $\pm$ 0.05 && 24.77 $\pm$ 0.08 & 23.73 $\pm$ 0.09 \\
\cutinhead{VEGA magnitudes}
$B$ & 25.29 $\pm$ 0.10 & 23.01 $\pm$ 0.10 && $>$ 28.6 & $>$ 26.3 \\
$V$ & 24.17 $\pm$ 0.08 & 22.35 $\pm$ 0.08 && 26.0 $\pm$ 0.2 & 24.2 $\pm$ 0.2 \\
$R$ & 23.45 $\pm$ 0.07 & 21.98 $\pm$ 0.07 && 24.6 $\pm$ 0.4 & 23.2 $\pm$ 0.4 \\
$I$ & 22.50 $\pm$ 0.06 & 21.41 $\pm$ 0.06 && 23.62 $\pm$ 0.18 & 22.58 $\pm$ 0.18 \\
$M_V$ & \nodata & $-$5.2 $\pm$ 0.2 && \nodata &$-$3.4 $\pm$ 0.3 \\
$ (\bv)_0 $ & \nodata & 0.66 $\pm$ 0.13 && \nodata &$>$  2.1 \\
$ (\vr)_0 $ & \nodata & 0.37 $\pm$ 0.11 && \nodata &1.0 $\pm$ 0.4 \\
$ (R-I)_0 $ & \nodata & 0.56 $\pm$ 0.10 && \nodata &0.6 $\pm$ 0.4 \\
\enddata
\tablecomments{$m$ is the observed magnitude and $m_0$ is the Galactic reddening corrected magnitude assuming $E(B-V)=0.558$. Magnitudes are calculated based on the spectrum of an F8~I supergiant for star A, and of an M5~III gaint for star B. }
\end{deluxetable*}

We used the {\tt geomap} task in the IRAF package IMMATCH to align the X-ray image to the mosaic F625W image by fitting the positions of the two sources in Table~\ref{tab:cp}, and then applying the best-fit transform parameters onto X-1 using the task {\tt geoxytran}.  This gives a best position of $\rm R.A.=03^h45^m55\fs612$, $\rm Decl.=+68\arcdeg04\arcmin55\farcs29$ (J2000) for X-1. The uncertainty in the fit using the two sources is $\delta_{\rm fit}=\sqrt{\delta_1^2 + \delta_2^2}/2=0.21\arcsec$, where $\delta_1$ and $\delta_2$ are the positional errors for sources 1 and 2 in Table~\ref{tab:cp}. Taking into account the error of X-1, we estimate the final 90\% error radius on the position to be $\sqrt{\delta_{\rm fit}^2 + \delta_{\rm X1}^2}= 0.21\arcsec$.  We note that a consistent position can be obtained for X-1 by matching source 1 only, but this yields comparable positional error because of the uncertainty in the rotation angle between the images. The F625W image at a region around X-1 is presented in Figure~\ref{fig:x1}. The uncorrected \cha\ error circle is displayed as a dashed circle, and the corrected position of X-1 with the error circle is shown as a small solid circle. 

\section{Optical Counterpart of X-1}
\label{sec:cp}

After the astrometry correction, a likely optical counterpart falls into the error circle of X-1, see Figure~\ref{fig:x1}.  The source is larger than the point spread function of the ACS with the F625W filter. Figure~\ref{fig:profile} shows the profiles of brightness along the X and Y pixel directions passing through the brightest pixel in the source for all broadband images.  It is obvious that there is an excess along the X profile which is asymmetric and more extended than the Y profile in all the images except the F435W image. This excess indicates that the source is either truly extended, a star with an associated extended source, or two stars.  We consider the last possibility most likely and label the bright core of the source lying well within the error circle of X-1 as ``star A'' and the other feature as ``star B'' in Figure~\ref{fig:x1}.  Star B is hardly seen in the F435W image, but becomes more and more significant at longer wavelength. It lies about $0.16\arcsec$ away from star A on the sky plane, just on the edge of the 90\% confidence error circle of X-1.

\section{Photometry}
\label{sec:photo}

Simple aperture photometry is not adequate to measure the magnitudes of stars A and B since they are not fully resolved.  Instead, we performed a simultaneous fit to the two sources with a point spread function, measured from bright isolated stars in the same image, to obtain their flux.  The fit was performed in the Interactive Data Language (IDL) using an algorithm adapted from the IRAF package DAOPHOT by W.\ B.\ Landsman. Star B is required for an adequate fit except for the F435W image.  The flux upper limit of star B in the F435W image was estimated from the total residuals around its position after star A was subtracted. Magnitudes are calculated from the flux measurements using the SYNPHOT package, assuming an F8~I supergiant spectrum for star A and an M5~III giant spectrum for star B, respectively.  The Galactic extinction along the line of sight to IC~342 is $E(B-V) = 0.558$, which is calculated from dust maps derived from {\it COBE} data \citep{sch98}.  The Galactic extinction used here is consistent with the total extinction to IC~342 of $E(B-V) = 0.63 \pm 0.10$ estimated by \citet{sah02}. \citet{gri06} estimated a local extinction of $E(B-V)=0.26$ from the $H_\alpha/H_\beta$ ratio in the nebula, indicating that most extinction is from our Galaxy.  \citet{fin07} also found that the majority of the extinction along the line of sight to IC~342 was Galactic. Therefore, we choose to not consider extragalactic extinction in calculating the reddening corrected colors.

Observed and reddening corrected ST magnitudes and VEGA magnitudes are listed in Table~\ref{tab:x1cp}.  To estimate systematic errors caused by different spectral types and band translation, we calculated magnitudes using spectra of different stars and added the errors linearly onto the results as systematic errors.  Spectral types used to estimate the systematic errors include F2~I, G0~I, and G2~I for star A, and all M~III for star B. The absolute V magnitudes are calculated assuming $(m-M)_0 = 27.58 \pm 0.18$ \citep{sah02}. Colors of $(B-V)_0$, $(V-R)_0$, and $(R-I)_0$ for both stars are also listed in Table~\ref{tab:x1cp}. Both star A and star B appear to have constant flux in the four F625W images, which were obtained over an interval of nearly 5 hours.  The $V$ magnitude and colors of star A are consistent with an F8 to G0 Ib supergiant, except for some excess brightness in the $I$ band.  The magnitude and colors of star B indicates it could be an M II giant; its red color suggests that it is similar to a carbon star \citep{men65}.

\begin{figure*}
\centering
\includegraphics[width=0.8\textwidth]{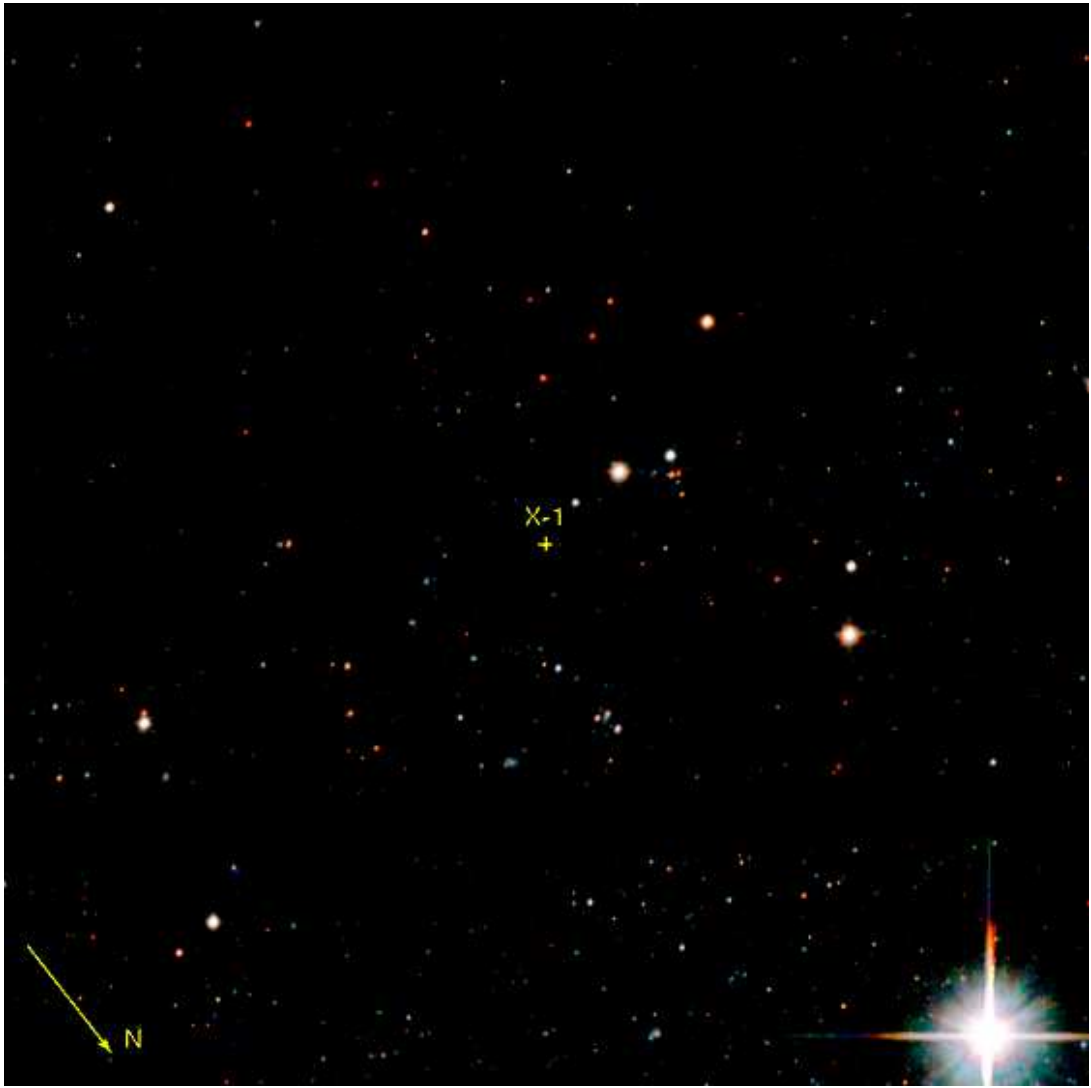}
\caption{
True color image of a $40\arcsec \times 40\arcsec$ region around IC~342 X-1. The black belt near the bottom is caused by the CCD gap. The F814W, F555W, and F435W images are used for the red, green, and blue channels, respectively. The arrow points north and has a length of $5\arcsec$.
\label{fig:20x0}}
\end{figure*}

\begin{figure}
\centering
\includegraphics[width=\columnwidth]{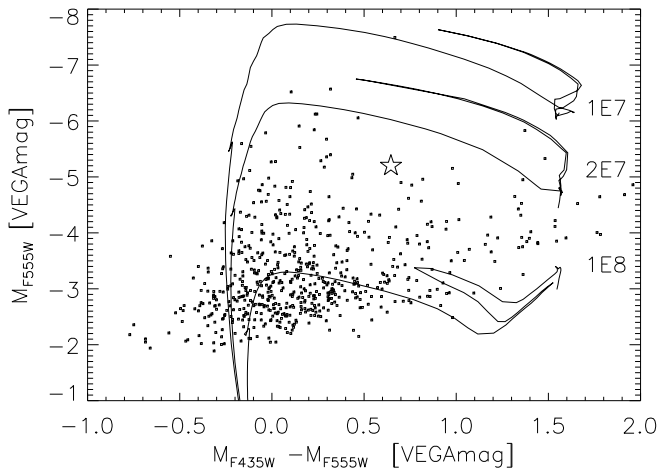}
\caption{
Color-magnitude diagram for stars shown in Figure~\ref{fig:20x0}. Star A is indicated with a {\it star} symbol. Isochrones of 10, 20, and 100 Myr stars are superimposed.  We note that the magnitudes are found in the \hst\ filter bands but quoted in the Vega magnitude system.
\label{fig:cmd}}
\end{figure}

A true color image of the region within $20\arcsec$ of X-1 is presented in Figure~\ref{fig:20x0}, composed of the F814W, F555W and F435W images, respectively, for the red, green and blue channels.  Using the IDL DAOPHOT and following the procedures described in \citet{kaa05}, we present a color-magnitude diagram of $M_{\rm F555W}$ versus $M_{\rm F435W}-M_{\rm F555W}$ in Figure~\ref{fig:cmd} for bright stars in the region that shows in Figure~\ref{fig:20x0}. To exclude extended sources, image defects and foreground stars, we selected sources that had peak values less than 0.75 and a local fit $\chi^2<3$. The magnitudes are extinction corrected but there might be additional reddening in the host galaxy or internal to the source, which is not considered. Typical errors are 0.2 for magnitudes and 0.16 for colors. Stars with absolute magnitudes brighter than $-8$ and also stars with colors bluer than $-0.4$ are likely foreground stars.  Isochrones for stellar populations with a metallicity of $Z=0.019$  at various ages (Girardi 2006, \url{http://pleiadi.pd.astro.it}) are also shown in Figure~\ref{fig:cmd}. 

\section{Multiwavelength Spectrum}
\label{sec:multi}

In the \cha\ observation, there are 0.465 counts per frame detected on a 3$\times$3 pixel island around X-1.  Following The \cha\ ABC Guide to Pileup\footnote{\url{http://asc.harvard.edu/ciao/download/doc/pileup\_abc.ps}},  this detected rate requires an input rate of 0.92 or 2.00 counts per frame assuming a grade migration probability $\alpha=0.5$, corresponding to a count lost of 49\% or 77\%, respectively.  Using PIMMS, we find that the observed fluxes from \asca\ and \xmm\ are roughly consistent with the detected count rate. However, the spectral shape and the luminosity state cannot be determined.

\begin{figure}
\centering
\includegraphics[width=\columnwidth]{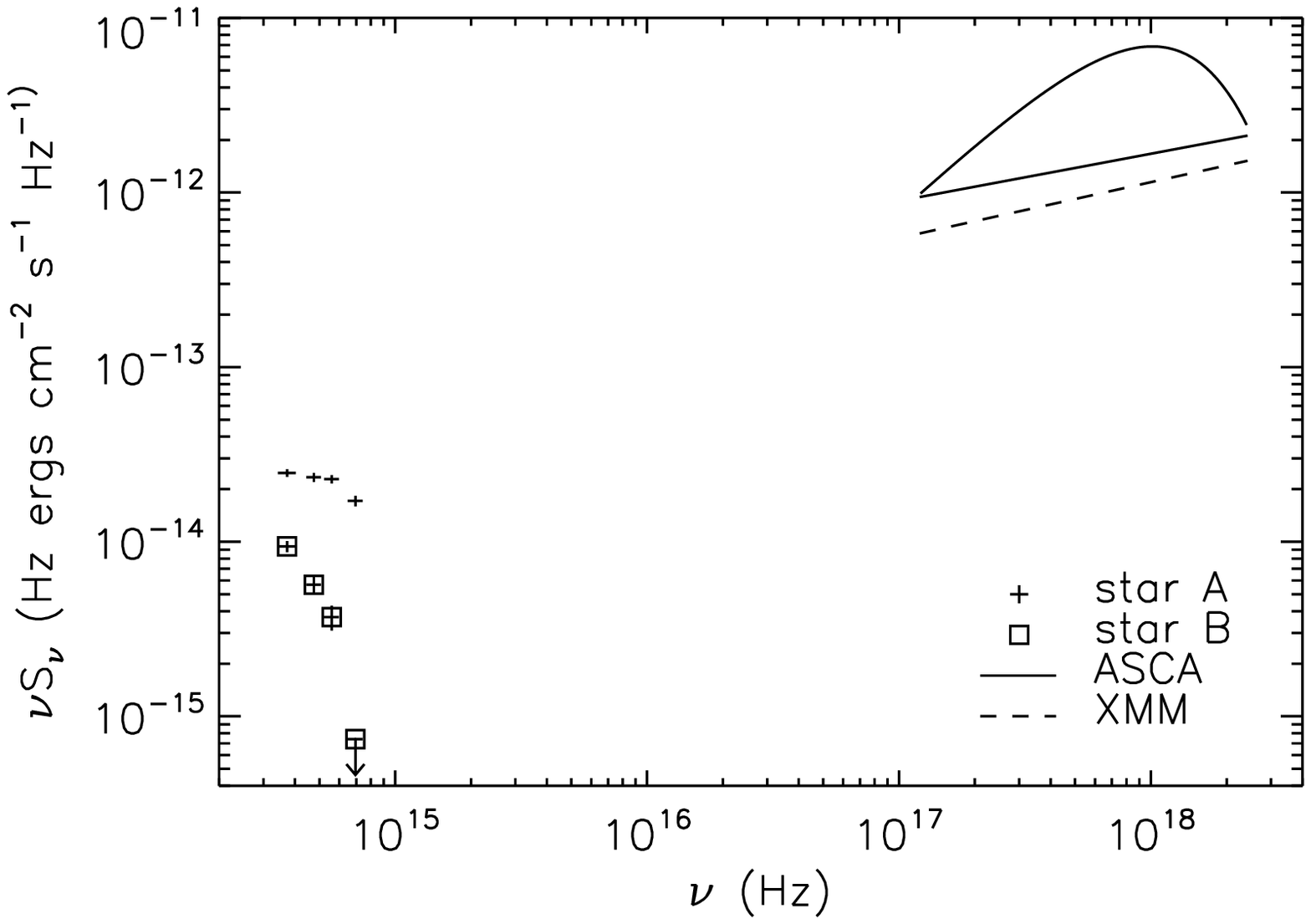}
\caption{
Multiwavelength spectrum of IC~342 X-1. {\it Corsses} and {\it squares} are reddening corrected optical fluxes for star A and B respectively, and {\it lines} are absorption corrected best-fit models for the X-ray spectra. The {\it solid lines} represent the two \asca\ spectra obtained from \citet{kub01} and the the {\it dashed line} is the \xmm\ spectrum from \citet{bau03}.  We note that the \asca\ spectra could be contaminated by surrounding sources by a factor of 20\% \citep{bau03}.
\label{fig:spec}}
\end{figure}

IC~342 X-1 is a highly variable source. \citet{kub01} reported that the source changed its X-ray spectral shape dramatically from 1993 to 2000 depending upon two \asca\ observations.  \xmm\ observed the source in 2001 and found the source spectrum was similar to that observed in the 2000 \asca\ observation.  A multiwavelength spectrum consisting of optical and X-ray data is presented in Figure~\ref{fig:spec}.  The optical spectra show a form of $S_\nu \propto \nu^{-1.51\pm0.12}$ for star A and $S_\nu \propto \nu^{-3.2\pm0.4}$ for star B.  We note that the \asca\ spectra could be contaminated by surrounding sources and the inferred flux could be overestimated by a factor of 20\% \citep{bau03}. 

The X-ray to optical flux ratio, defined as $\log(f_{\rm X}/f_V)=\log f_{\rm X} + m_V/2.5 + 5.37$, where $f_{\rm X}$ is the 0.3--3.5 keV flux in \ergcms\ and $m_V$ is the visual magnitude \citep{mac82}, is found to be 3.0--3.7 assuming star A is the counterpart or 3.8--4.5 for star B, given different X-ray spectra and the \hst\ $m_V$.  This is larger than the ratios for normal stars, normal galaxies, clusters of galaxies, AGNs, and BL Lac objects found in the {\it Einstein} Extended Medium Sensitivity Survey \citep{mac88,sto91}.  Similarly, the two-point optical to X-ray spectral index computed between the $V$ band and 1 keV is $\alpha_{\rm OX}=0.2-0.4$ for star A and $\alpha_{\rm OX}=-0.05-0.14$ for star B, which are flatter than the values found in those objects \citep{sto91,lan01}.

To compare with Galactic X-ray binaries, we computed the X-ray to optical flux ratio defined in the form $\xi = B_0 + 2.5 \log F_{\rm X}$, where $B_0$ is the reddening corrected $B$ magnitude and $F_{\rm X}$ is the 2--10 keV observed X-ray flux in $\mu$Jy.  We obtain $\xi=20.4-22.1$ for star A and X-1 using its X-ray spectra at different emission states.  Since star B is not detected in the $B$ band with these \hst\ observations, we estimate the $\xi$ for star B is greater than 24.  These numbers are greater than that from most high-mass X-ray binaries, but well consistent with the range of $\xi$ found in low-mass X-ray binaries \citep{van95}.  

The multiwavelength spectrum of X-1, no matter whether star A or star B is assumed to be its optical counterpart, is unlike the spectra of quasars or blazars.  The spectrum is consistent with an X-ray binary in IC 342 X-1.

\section{Line emission from the nebula and X-1}
\label{sec:halpha}

\begin{figure*}
\centering
\includegraphics[width=0.8\textwidth]{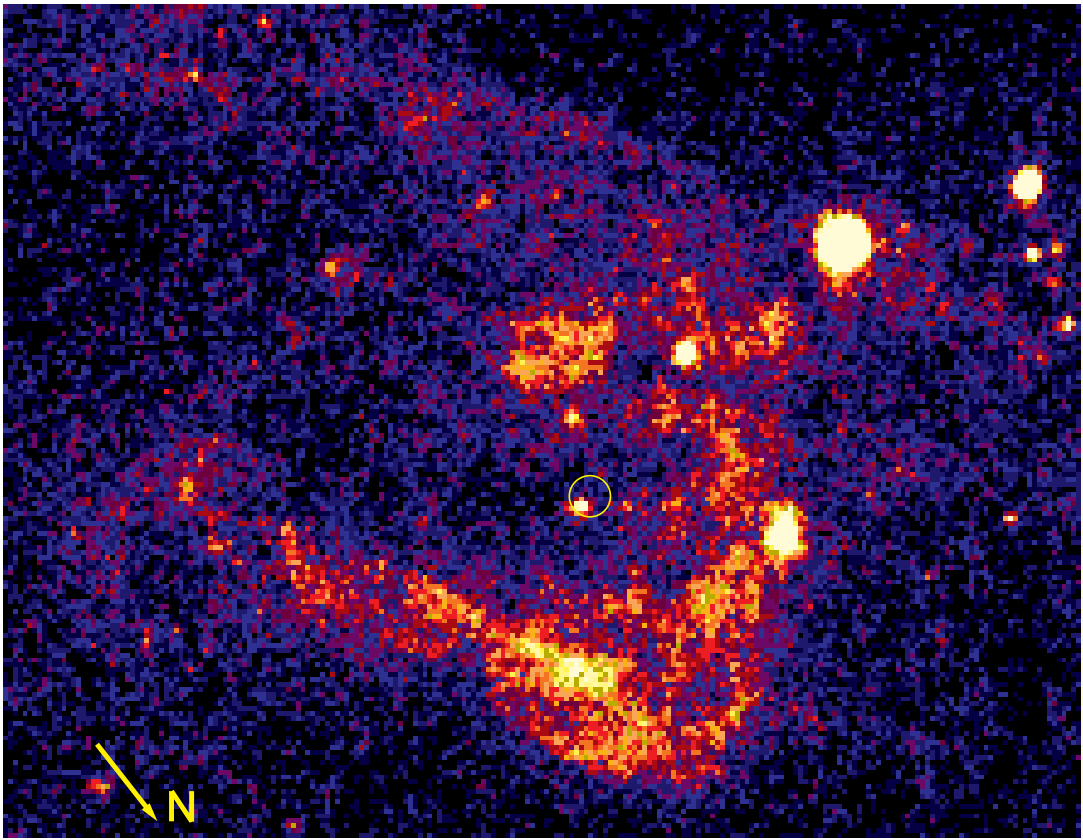}
\caption{
F658N image of the region around IC 342 X-1.  This filter includes lines
H$\alpha$ $\lambda$6563, [N~{\sc ii}] $\lambda$6583, and [N~{\sc ii}] $\lambda$6548.  The error circle for X-1 is shown.  The arrow points north and has a length of 1\arcsec.
\label{fig:ha}}
\end{figure*}

Figure~\ref{fig:ha} shows the F658N image in the region around IC 342 X-1.  This filter includes lines from [N~{\sc ii}] at $\lambda$6583 and $\lambda$6548 as well as H$\alpha$ at $\lambda$6563, and their relative transparency is 1.0, 0.5 and 1.0, respectively.  The error circle of X-1 is shown and the optical counterpart of X-1, discussed above, is clearly visible in the image.  We note that the optical counterpart of IC 342 X-1 lies very close to the geometric center of the emission line nebula.  This strengthens the association between the nebula and the X-ray source and suggests that the X-ray source has not moved significantly since the creation of the nebula.

The nebula has an angular scale of around 6\arcsec\ in diameter,
corresponding to 100~pc at a distance of 3.3~Mpc. We estimated the total flux of the nebula using an irregular shaped region encompassing the regions bright in the F658N image but with stars apparent in the continuum image (F625W) removed.  We used the F625W image to estimate the continuum.  Since the line and continuum are shown in both filters, we modeled the emission as a flat continuum plus line emission and solved algebraically for the continuum and line strength given the count rates in the two filters.  We estimate that the reddening corrected line flux from the nebula is $(4.7 \pm 0.3) \times 10^{-14}$~\ergcms. This flux is consistent with the sum of the H$\alpha$ and [N~{\sc ii}] fluxes found by \citet{abo07}.  We note that this flux is significantly higher than that quoted by \citet{rob03}, but we find that our flux before the reddening correction is consistent with that quoted by \citet{rob03} as reddening corrected.  Using the line ratios in \citet{abo07} and \citet{rob03}, we estimate that the H$\alpha$ $\lambda$6563 flux from the nebula is about $(2.5 \pm 0.3) \times 10^{-14}$~\ergcms.  Using the same continuum subtraction procedure, we do not detect any H$\alpha$ line emission from the optical counterpart of X-1 and place an upper limit on the reddening corrected flux of $1.0 \times 10^{-16}$~\ergcms.  

\section{Discussion}
\label{sec:diss}

The stellar population within 300~pc ($20\arcsec$) of X-1 appears to contain no O stars.  The color-magnitude diagram (Figure~\ref{fig:cmd}), suggests that the minimum stellar age is $\sim$10~Myr if the metallicity is close to $Z=0.019$.  Comparing to isochrones at a lower metallicity of $Z=0.004$, the minimum stellar age is about 20~Myr. The region appears to contain relatively young stars with ages of tens of Myr and also an older population with ages of several Gyr.

IC 342 X-1 appears at the center of the nebula (Figure~\ref{fig:ha}), which strongly suggests an association. The relatively old age of the stellar population in the vicinity of the nebula makes it unlikely that the nebula is powered (even in part) by ionization from one or more O stars.  \citet{abo07} suggested that the shock velocity in the nebula is 20 to 100 km/s, implying the nebula is younger than 2~Myr.  This age is an overestimate since the expansion would be slowing down.  This suggests that if the nebula was created in a supernova event, the progenitor star must has a lifetime close to 10~Myr.  The high luminosity of X-1 suggests it contains a black hole.  To form a black hole in an energetic explosion, the progenitor must be very massive \citep[e.g., at least 40~$M_\sun$ in][]{heg03} and thus has a lifetime significantly shorter than 10~Myr.  Therefore, it is unlikely that the nebula was powered by an unusually powerful explosion in which a black hole was formed.

A more plausible alternative is that the nebula is powered by an outflow from the accreting compact object in the ULX \citep{pak02,rob03}.  The nebula appears to be predominantly shock powered with a shock velocity in the range 20--100~km/s and a total power of $3 \times 10^{39}$~\ergs, but with high excitation lines possibly powered by photoionization near the center \citep{abo07}.  This power is 20\% to 50\% of X-ray luminosity, depending on the X-ray state \citep{oka98}, and is reasonable in that X-ray binaries appear to have power in their outflows which is comparable to or may even exceed their X-ray emission \citep{gal05}.  The emission line morphology in Figure~\ref{fig:ha} indicates that the nebula is not a complete shell.  The morphology is somewhat similar to that of the W50/SS433 system \citep{bou07} and may suggest a jet orientation along an axis slightly west of north.  Given the strong radio emission from W50/SS433, radio observations of IC 342 X-1 would be of great interest.

We find one optical source in the error circle of IC 342 X-1.  The source appears extended at long wavelengths and is either two stars or a star with an associated feature such as a jet.  We consider the relativistic jet interpretation unlikely because their spectra are quite steep, unlike the optical relativistic jets of black hole such as those of M87 \citep{per01}.  The projected separation in IC 342 would be 2.6~pc, and the true separation may be larger.  This is too large to be an optical compact jet found in the hard state of stellar mass black holes, which usually has a spatial scale of AU \citep{fen06}.  We find interpretation in terms of two superimposed stars more likely.  The brighter star (star A) lies well within the error circle for IC 342 X-1.  The dimmer star (star B) lies at the edge of the 90\% confidence error circle of X-1 and could, instead, be the counterpart of X-1. Allowing either star as the counterpart, the X-ray to optical flux ratio indicates that the source is not a background blazar or quasar, but is consistent with an X-ray binary in IC 342.

If the companion star dominates the optical emission despite the high X-ray flux, then the spectral types of the stars can be determined.  The colors and absolute magnitude of star A are consistent with classification as an F8 to G0 Ib supergiant, which has a mass of about 10~$M_\sun$.  There is some excess $I$ band emission, but this is often seen in supergiants and is a signature of a circumstellar disk or a stellar wind.  With this classification, the age of star A would be roughly 50~Myr.  The evolutionary status of star A, on the supergiant branch, would be compatible with the large mass transfer rates needed to power the inferred luminosity of the ULX.

The colors and absolute magnitude of star B are consistent with classification as a carbon star.  The progenitor mass of such a star is likely of intermediate mass, $1-5 M_{\odot}$, and, thus, would require a rather old binary system.  The carbon star phase is short-lived, $\lesssim 1$~Myr.  High luminosity carbon stars have high mass loss rates \citep{jur91} which would be sufficient to power the ULX. 

However, the optical emission could also be dominated by reprocessed X-rays on the companion star or on the accretion disk, or by the intrinsic emission of the disk.  In these cases, spectral classification of the companion star via its colors is not reliable.  We discuss these possibilities in following.

\begin{figure}[h]
\centering
\includegraphics[width=\columnwidth]{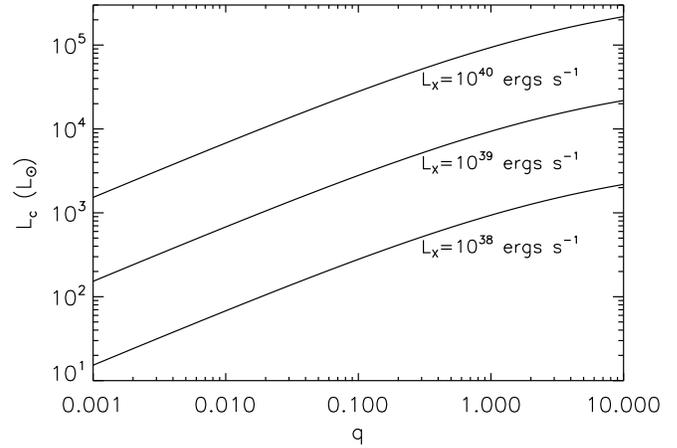}
\caption{
Companion star intercepted luminosity ($L_{\rm c}$ in unit of solar luminosity) as a function of the binary mass ratio ($q$, the companion to compact object mass ratio), assuming isotropic emission and Roche-lobe overflowing. Different lines are scaled with different X-ray luminosity as labeled.
\label{fig:lc}}
\end{figure}

Assuming the X-ray emission is isotropic, the X-ray luminosity intercepted by the companion is
\begin{equation}
L_{\rm c} \approx \frac{\pi R_2^2}{4\pi a^2}L_{\rm X} \; ,
\end{equation}
where $R_2$ is the radius of the companion star, $a$ is the binary separation, and $L_{\rm X}$ is the total X-ray luminosity.  The high luminosity of ULXs indicate that they accrete via Roche-lobe overflow.  Substituting equation~(2) in \citet{egg83} in which $R_2/a$ is determined by the companion to compact object mass ratio $q$ to an accuracy better than 1\% for any $q$, we have
\begin{equation}
L_{\rm c} = \frac{L_{\rm X}}{4} \left[ \frac{0.49q^{2/3}}{0.6q^{2/3} + \ln(1+q^{1/3})} \right]^2 \; .
\end{equation}
A plot of $L_{\rm c}$ versus $q$ at different X-ray luminosities is presented in Figure~\ref{fig:lc}. We can see that for a source like IC~342 X-1, which has $L_{\rm X} \approx 10^{40}$~\ergs, the companion intercepted X-ray luminosity is above $10^3$~$L_\sun$ when $q>0.001$ and above $10^4$~$L_\sun$ when $q>0.01$.  These luminosities are as high as of massive main sequence stars, bright giants or supergiants, implying that the optical counterpart of a bright ULX will appear as bright as a massive star or giant/supergiant no matter what the companion really is.  The observed optical spectra suggest star A has a luminosity of about $10^4$~$L_\sun$, and star B has a luminosity of about $10^3$ to $10^4$~$L_\sun$. Thus, the optical emission may well be dominated by an X-ray ``heated'' companion star.  However, how much of the companion surface is shielded by the accretion disk is unknown.  For ULXs with $L_{\rm X}>10^{40}$~\ergs, if even a small portion ($>10\%$) of the X-ray flux is not absorbed enroute to the companion surface, then the companion's temperature and luminosity will be notably raised.

Disk irradiation may be an important component of the optical emission if the X-ray luminosity is truly very high.  Gris\'e et al.\ (2007, in preparation) found significant variability from the optical counterpart of NGC~1313 X-2 on time scales from days to hours and even minutes, suggesting the optical emission is dominated by X-ray reprocessing.  Both star A and star B show no detectable variability in the four F625W images at time scales of a few hours. The X-ray to optical flux ratio taking either star A or B as the counterpart is consistent with those found in neutron-star and stellar mass black hole X-ray binaries (the low mass X-ray binaries; LMXBs), where the emission is thought to be dominated by reprocessed X-rays on the disk \citep{van95,liu01}.  However, the color of either star is significantly redder than the colors, $(B-V)_0 \sim 0$, typical of LMXBs.  Assuming a standard disk geometry, the disk temperature from X-ray irradiation varies as $T(R)^4 \propto L_{\rm X}/R^2$ at different radius, and the spectral energy distribution has a form of $S_\nu \propto \nu^{-1}$ by integrating to infinite radius.  Therefore, for a finite disk, the X-ray reprocessing at the outermost radius ($R_{\rm out}$) determines the peak of the energy distribution, which could be attributed to a simplified blackbody emission with an effective temperature $T_{\rm eff}$.  We assume the disk size is scaled with the binary separation $a$, so the effective temperature of the heated disk varies as  $T_{\rm eff}^4 \propto T(R_{\rm out})^4 \propto L_{\rm X}/a^2$.  If the system transfers mass via Roche-lobe overflow, then $a \propto M^{1/3} P^{2/3}$, where $M$ is the compact object mass, $P$ is the orbital period and we have assumed that the companion mass is less than that of the compact object, also the companion density varies as $\rho \propto P^{-2}$.  Thus, we find $T_{\rm eff} \propto L_{\rm X}^{1/4} M^{-1/6} \rho^{1/6}$.  For most Galactic LMXBs, the companion is a main sequence star with a density in order of $1 \rm \, g \, cm^{-3}$ and the $T_{\rm eff}$ of the heated disk is around 25,000--30,000~K, or $(B-V)_0 \sim 0$.  To produce a heated disk with a $T_{\rm eff}$ around 5,500--6,000~K as star A at a high X-ray luminosity level, it requires either an IMBH or an well evolved companion that has a low density.  If the optical counterpart is star B, whose optical spectral index is much steeper than $-1$, it is unlikely that the optical emission is dominated by disk irradiation.

Parenthetically, we note that the relation found by \citet{van94} between the visual luminosity $L_V$, X-ray luminosity, and orbital period of LMXBs has been used to make inferences about the orbital periods of ULXs.  This relation was based mainly on sources with relatively dense companions, and the correlation is inferred based on the assumption that $L_V \propto T^2$, which is no longer correct when $T<10,000$~K.  Therefore, one should be cautious in applying this correlation in systems where the observed $T_{\rm eff}$ is lower than 25,000--30,000~K, or $(B-V)_0$ is higher than 0.

The intrinsic disk emission in the optical band should present in a form of $S_\nu \propto \nu^{1/3}$, which is totally different with the observed optical spectra.  Therefore, we suggest that the optical emission is not dominated by the viscous disk.

In sum, for a ULX system with $L_{\rm X}>10^{40}$~\ergs, it appears unlikely that the intrinsic emission of the companion star dominates the optical emission. The companion star is likely significantly heated by X-ray illumination.  Disk irradiation could also be important. The color of the optical counterpart is redder than typically found in systems, such as LMXBs, where a significant fraction of the optical light arises from an X-ray heated disk. The red color could be a consequence of a massive black hole or a low density companion.  

\acknowledgments 
We thank the anonymous referee for insightful comments, which have improved the paper especially in the discussion section. HF thanks Manfred Pakull for helpful discussions.

\end{document}